\documentclass[a4paper,11pt]{article}
\usepackage{pos}
\usepackage{todonotes}

\title{Search for STaus in IceCube}
 \ShortTitle{Stau Search in IceCube}

\author{The IceCube Collaboration \\{\normalsize \normalfont(a complete list of authors can be found at the end of the proceedings)}}

\emailAdd{jh.schmidt-dencker@tum.de}
\emailAdd{stephan.meighen-berger@tum.de}
\emailAdd{christian.haack@tum.de}

\abstract{The tau lepton’s supersymmetric partner, the stau, appears in some models as the next-to-lightest supersymmetric particle.
Their decay process into the lightest superpartner is usually suppressed by supersymmetry breaking, which makes it a long-lived particle.
In this scenario, its signature is a long, minimally ionizing track when traveling through the IceCube detector.
Independent of their primary energy, the stau tracks appear like low-energy muons in the detector.
A potential signal of staus would thus be an excess over muon tracks induced by atmospheric muon neutrinos.
Our analysis focuses on the region around the horizon as here the ratio between stau signal and atmospheric background is largest.
We will present the first sensitivity to constrain the stau mass using IceCube and demonstrate the potential of this analysis with future improvements. 

\vspace{4mm}
{\bfseries Corresponding authors:}
\newline
Jan-Henrik Schmidt-Dencker$^{1*}$, Stephan Meighen-Berger$^{1}$, Christian Haack$^{1}$\\
{$^{1}$ \itshape Technische Universität München}\\
$^*$ Presenter

\FullConference{37$^{\rm{th}}$ International Cosmic Ray Conference (ICRC 2021)\\
		July 12th -- 23rd, 2021\\
		Online -- Berlin, Germany}

}

\begin{document}
\maketitle

\section{Introduction}
The supersymmetric principle (SUSY) introduces a new symmetry, linking fermions and bosons, which facilitates numerous extensions of the Standard Model (SM) \cite{MARTIN_1998}.
As a result of this additional symmetry, each SM particle obtains a SUSY counterpart.
This principle is formulated with the concept of R-parity, where every SM particle is assigned an even parity and every superpartner receives odd parity \cite{Fayet}.

Provided that R-parity is conserved, the lightest supersymmetric particle (LSP) is stable.
While most studies assume this particle to be a neutralino, extensions to include gravity suggest the gravitino as an alternative LSP candidate \cite{Ahlers}.
In scenarios where the gravitino is a stable LSP and the stau represents the next-to-lightest supersymmetric particle (NLSP), the decay of the stau into a gravitino and a tau lepton is suppressed by the scale of SUSY breaking \cite{StaufavouredOverNeutralinoFeng_2004, Meighen-Berger:2020eun}.
Hence, the stau appears as a long-lived particle and can be measured in collider experiments.

However, the production of staus is not restricted to collider experiments.
High-energy cosmic rays striking the atmosphere provide enough energy for the production of SUSY particles.
If the stau is sufficiently long-lived, large volume neutrino telescopes are able to detect its Cherenkov signature (see \cite{Albuquerque}).

As signatures of stau signals in the detector are hardly distinguishable from muons, a characteristic indicator is necessary to claim a discovery.
Previous efforts concerning a stau detection with neutrino telescopes focused on a search for double track signatures \cite{Ahlers,Kopper}.
While only few SM interactions result in parallel tracks of muons, staus are produced in pairs and moving in parallel tracks continuously through Lorentz boosting.
This makes the detection of an excess of parallel tracks a distinctive signature for staus \cite{Albuquerque,Ahlers}.

Complementary to double track searches, \cite{Meighen-Berger:2020eun} proposes a discovery potential of an excess stau event search over the low-energy muon background.
Following this novel approach, our analysis, as the very first of its kind, investigates the suggested region of interest using the IceCube Neutrino Observatory \cite{IceCube}.
In this article, we state the details of our Monte Carlo (MC) analysis and present our first results as sensitivities to the stau mass.

\section{Analysis Structure}
To study and discover stau signals in IceCube, we rely on precise modelling of signal and background distributions.
As a first feasibility study, we use an existing event selection event selection for low-energy muons from atmospheric muon neutrino interactions, the main background of our analysis.
Details on the calculation of the used background can be found in \cite{Stettner}.

For now, an already existing background simulation allows us to concentrate on the stau signal inside IceCube.
The simulation chain for our signal can be structured into three overall categories: Production, Propagation and Detection Simulation.
We begin by explaining the propagation of staus in comparison to muons and its significance for the analysis.
Afterwards we focus on the production of staus in the atmosphere and show the resulting stau fluxes, after they were propagated to the detector surface.
As a final step of our simulation, we model the detectors response to staus and calculate effective areas.

\paragraph{Propagation}
Like muons, the stau is subject to an energy loss $dE$ when traversing a distance $dX$ in a material according to 
\begin{equation}
    -\dfrac{dE}{dX} = \alpha(E) + \beta(E)E.
    \label{eq:eloss}
\end{equation}
While the ionisation losses, represented by $\alpha(E)$, are the same for muon and stau, the stochastic losses ($\beta(E)$) of the stau are suppressed by its mass.
This difference in energy loss plays a crucial role, as it is what makes our analysis possible in the first place.

\begin{figure}[h]
    \centering
    \includegraphics[width = .8\textwidth]{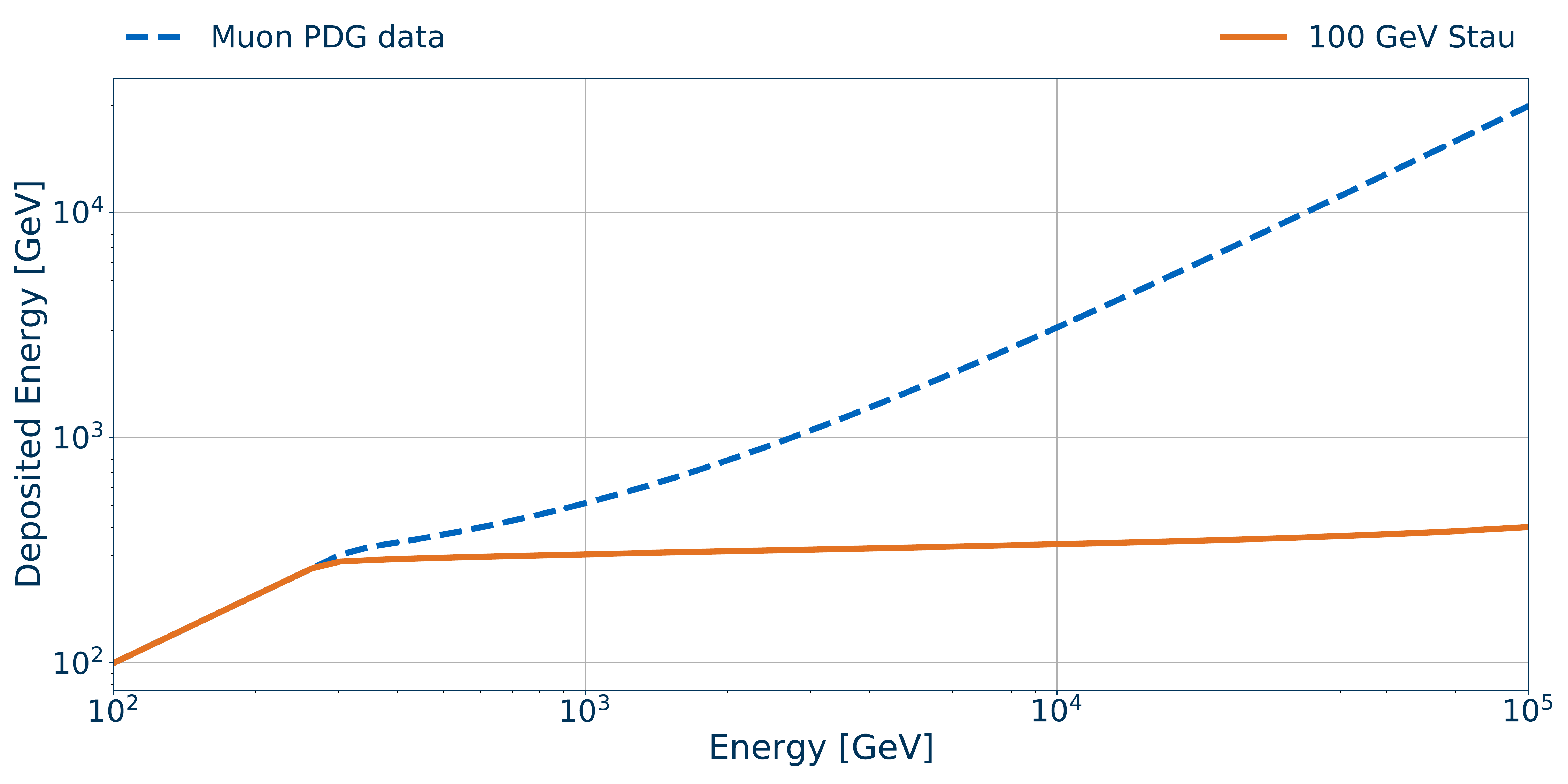}
    \caption{Energy deposit of staus (orange) and muons (blue) for a propagation length of 1 km in ice.}
    \label{fig:EDeposit}
\end{figure}

As a result of the suppression, the energy deposited by staus remains constant for any initial stau energy (\autoref{fig:EDeposit}), generating an excess signal in the low energy region.
Additionally, this property enables staus to propagate further through any material than muons.
This combination opens up the possibility of an excess event search in a region of zenith angles where muons from atmospheric air showers can no longer reach the detector and muons induced by atmospheric neutrinos are not yet present in numbers.

\paragraph{Production}
Both, signal and background events of our analysis originate from cosmic-ray air showers.
The Drell-Yan cross section for the production of staus at proton beam collisions is calculated with MadGraph \cite{MadGraphAlwall_2014}.

\begin{figure}[h]
    \centering
    \includegraphics[width = .8\textwidth]{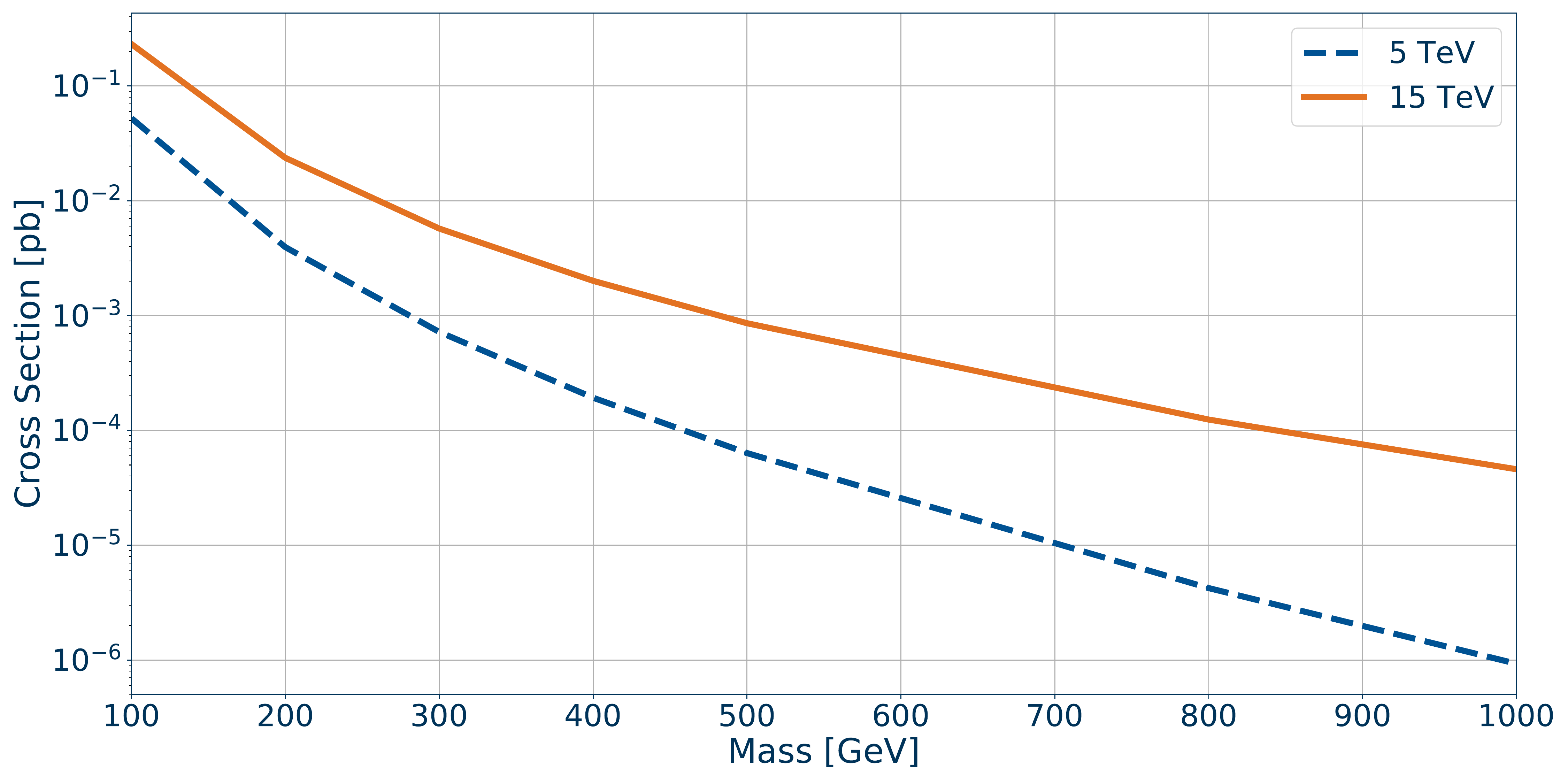}
    \caption{MadGraph cross section data for stau production in 5TeV and 15TeV proton beam collisions. The cross section decreases for larger stau masses resulting in a smaller stau flux.}
    \label{fig:crosssection}
\end{figure}

In \autoref{fig:crosssection} we see a decrease of the cross section depending on the mass of the stau.
Higher beam energies and thus higher cosmic-ray primaries however, increase the probability of stau production.
Via the Glauber formalism \cite{Glauber}, we generalize the cross section data from pure proton beam collisions in \autoref{fig:crosssection} to be valid for collisions of protons with air molecules.

Staus are more likely to be produced in the primary interaction of a cosmic-rays with the atmosphere.
Nevertheless, we monitor the proton flux towards Earth with MCEq (\cite{MCEq}) and apply the stau cross section at each level of the air shower.
As we are merely interested in the average individual particle fluxes, we benefit from an application of matrix cascade equations over elaborate MC shower simulations.
Note, that the transverse dispersion of air showers, accounted for in MC simulations, is sub-dominant at energies above a few GeV \cite{HuberThesis}.
Hence, so-called 3D-effects are negligible for primary energies necessary to produce stau pairs.

\begin{figure}[h]
    \centering
    \includegraphics[width = .8\textwidth]{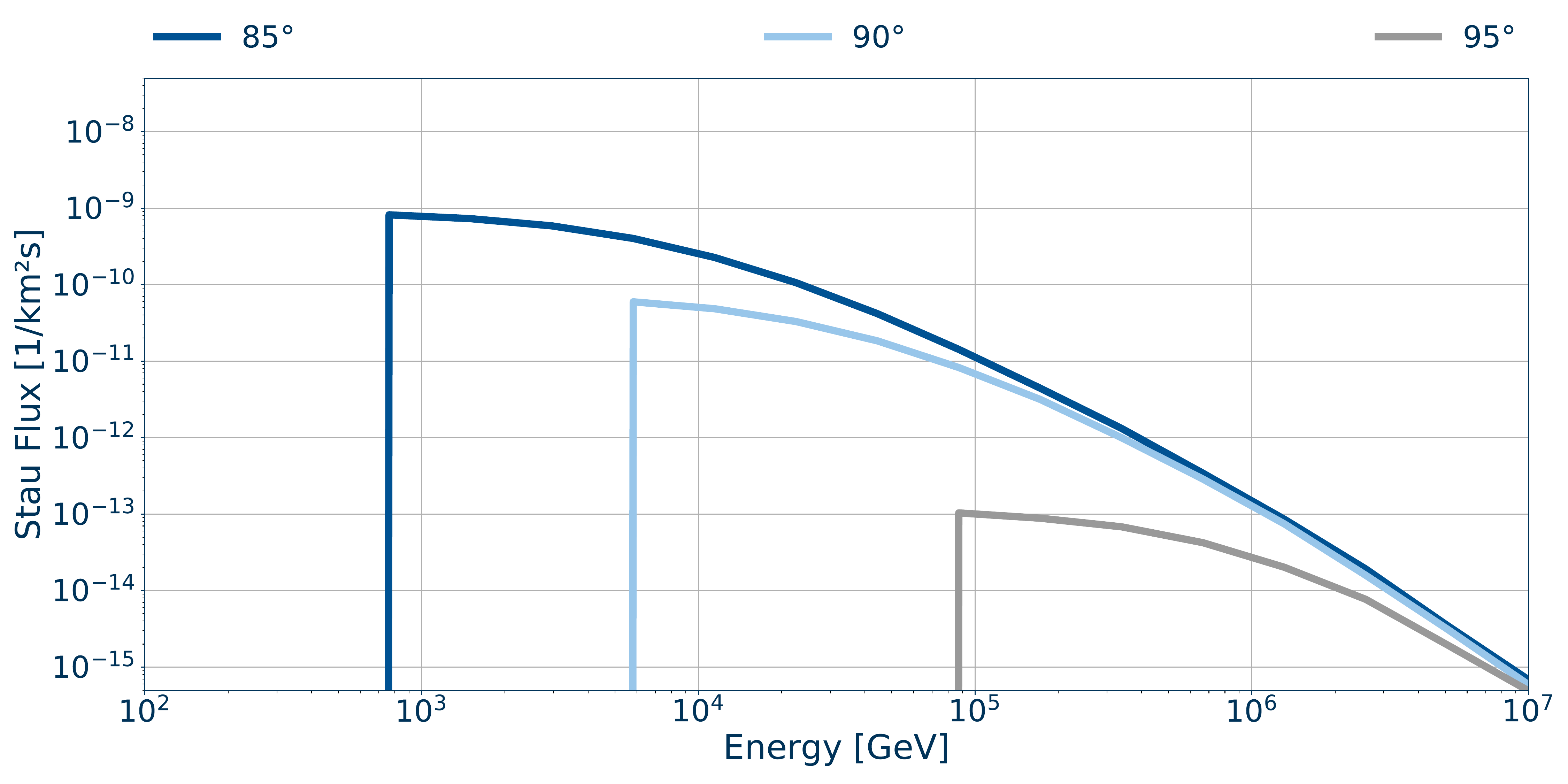}
    \caption{Simulated flux of 100 GeV staus at the detector surface for three different zenith angles (0° = true south).}
    \label{fig:stauflux}
\end{figure}

In the end, the combination of the production in air showers and the propagation to the detector yield the stau flux at the detector surface seen in \autoref{fig:stauflux}.
Notice the sharp edge that is a result of our current requirement to solely include relativistic staus.

During the propagation of the stau towards the detector, we neglect the magnetic bending of the particle as well as multiple scattering.
A small non-vertical magnetic field component might affect the zenith angle distribution and thus have an influence on the stau flux.
Similar to the magnetic bending, multiple scattering processes described in \cite{AlbuquerqueKlein:2009vk} can broaden the angular range and shift events to higher zenith angles.
The influence on the stau flux from either effect and whether it is negligible will be assessed during future analysis steps.

\paragraph{Detector Simulation}
Our simulation of the detector response to staus uses IceCube's internal MC software.
We use PROPOSAL \cite{PROPOSAL} to calculate the stau energy losses and propagate the photons with the CLSim package \cite{CLSim}.
The simulation includes 20 million stau events for every zenith angle between 0° and 180° in an energy range from 100 GeV to 10 TeV.

After the simulated detector response we use an existing event selection optimized for track detection from the northern hemisphere (zenith angles > 90°).
Details on this event selection can be found in \cite{Radel:2017ule} where it originates form, with optimizations from \cite{Stettner, HaackThesis}.

\begin{figure}[h]
    \centering
    \includegraphics[width = .8\textwidth]{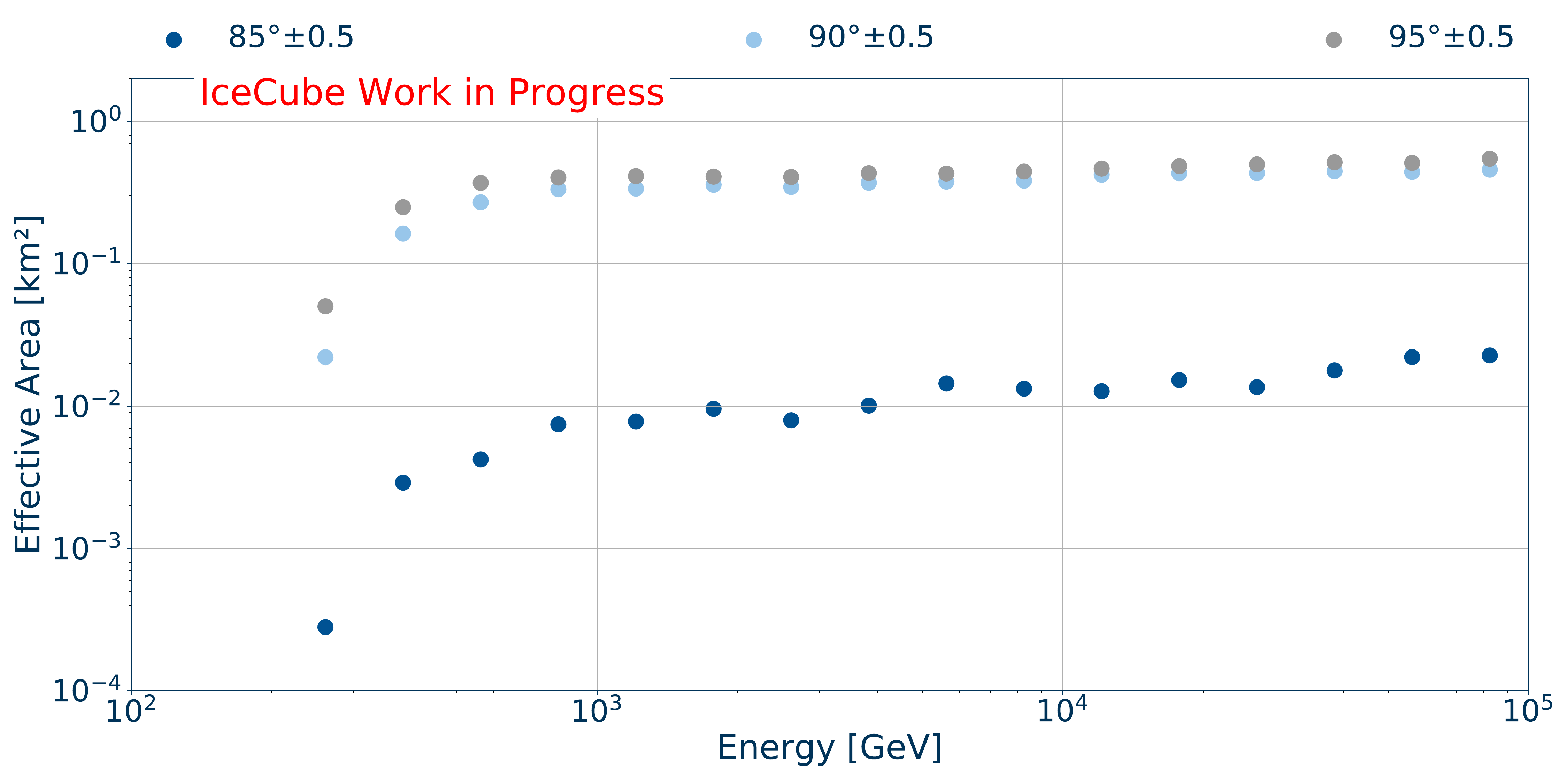}
    \caption{Effective area for stau events in three selected zenith angles.}
    \label{fig:effectiveArea}
\end{figure}

The simulated MC events pass through the different filter levels of the event selection.
Remaining events are binned in zenith angle and energy to calculate an effective area.
The effective area of stau events in three different zenith angles is displayed in \autoref{fig:effectiveArea}.

\paragraph{Background}
The background of our analysis consists of secondary muons induced by interactions of muon neutrinos from cosmic-ray air showers.
The track selection incorporates a strong cut-off below 85° to shield atmospheric muons and thus cuts out part of our signal too.
Although it does not cover our full region of interest, we are provided with a solid estimate to compare with our stau signal.

\section{Results}
\begin{figure}[h]
    \centering
    \includegraphics[width = .8\textwidth]{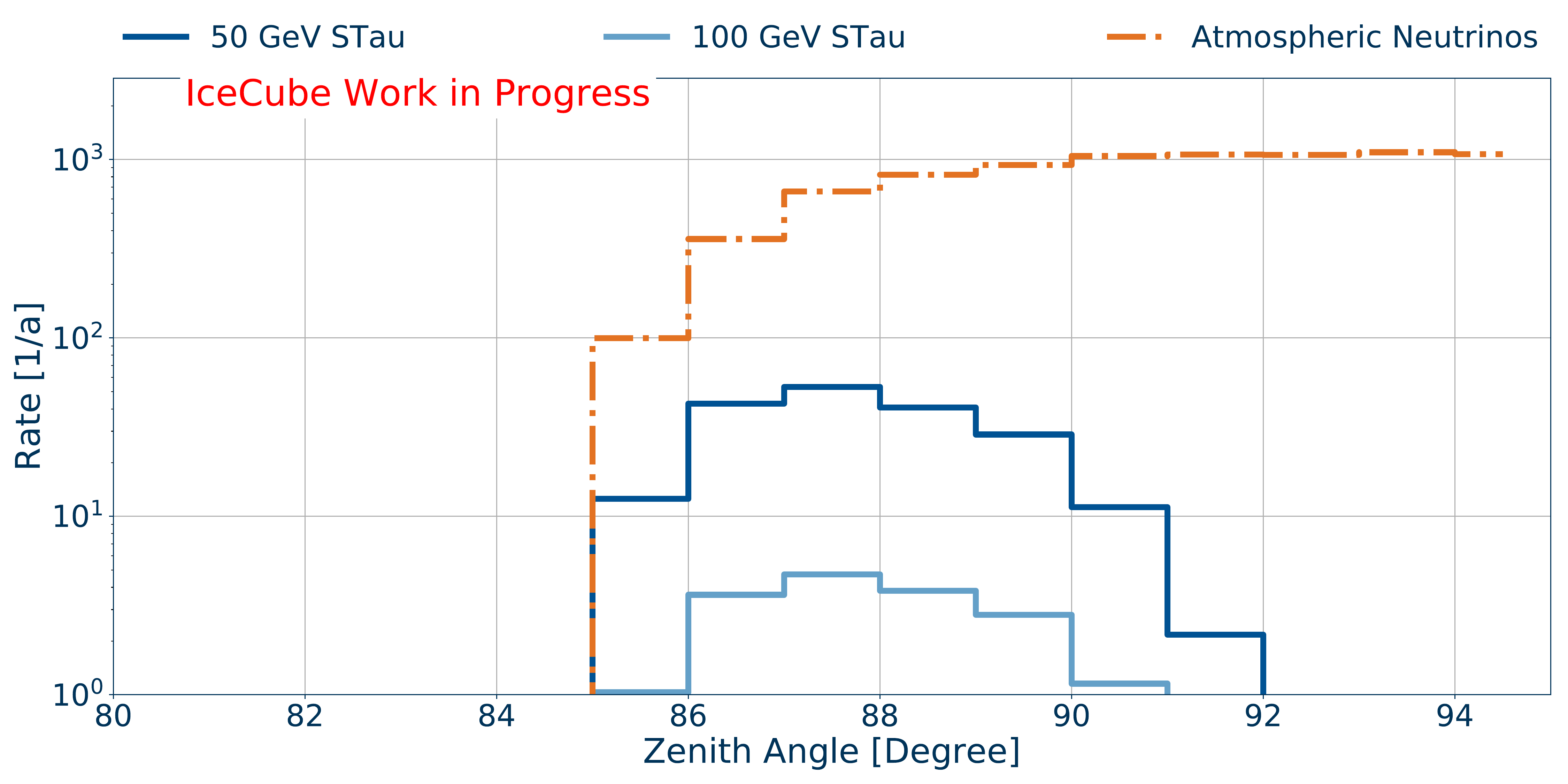}
    \caption{Signal and background event rates.
    Displayed are the rates for staus with masses of 100 Gev and 150 GeV respectively.
    The simulated background results from a event selection optimized for northern tracks \cite{Radel:2017ule, Stettner, HaackThesis}}
    \label{fig:rates}
\end{figure}

The combination of our simulated flux (\autoref{fig:stauflux} and the effective area for staus in IceCube (\autoref{fig:effectiveArea}) yields an estimate on the stau event rates.
In \autoref{fig:rates}, we show the stau event rates for two different masses alongside the muon background integrated over energies from 100 to 1000GeV  and dependending on the zenith angle.
We see an expected cut-off at 85° as a result of our event selection and a strong dependence between stau rates and mass.
Nevertheless we see a promising signal to background ratio at lower zenith angles.

Our overall signal and background rates ($R_{Signal},\; R_{Background}$) are available in form of a zenith/ energy binned grid.
This provides us with sufficient information to generate maps containing $\chi^2$ values for each bin.
Our $\chi^2$ is thereby calculated with the following formula: 

\begin{equation}
\chi^2 = \frac{R_{Signal}^2}{R_{Background}}.
\label{eq:chi2}
\end{equation}

\begin{figure}[h]
    \centering
    \includegraphics[width=.8\textwidth]{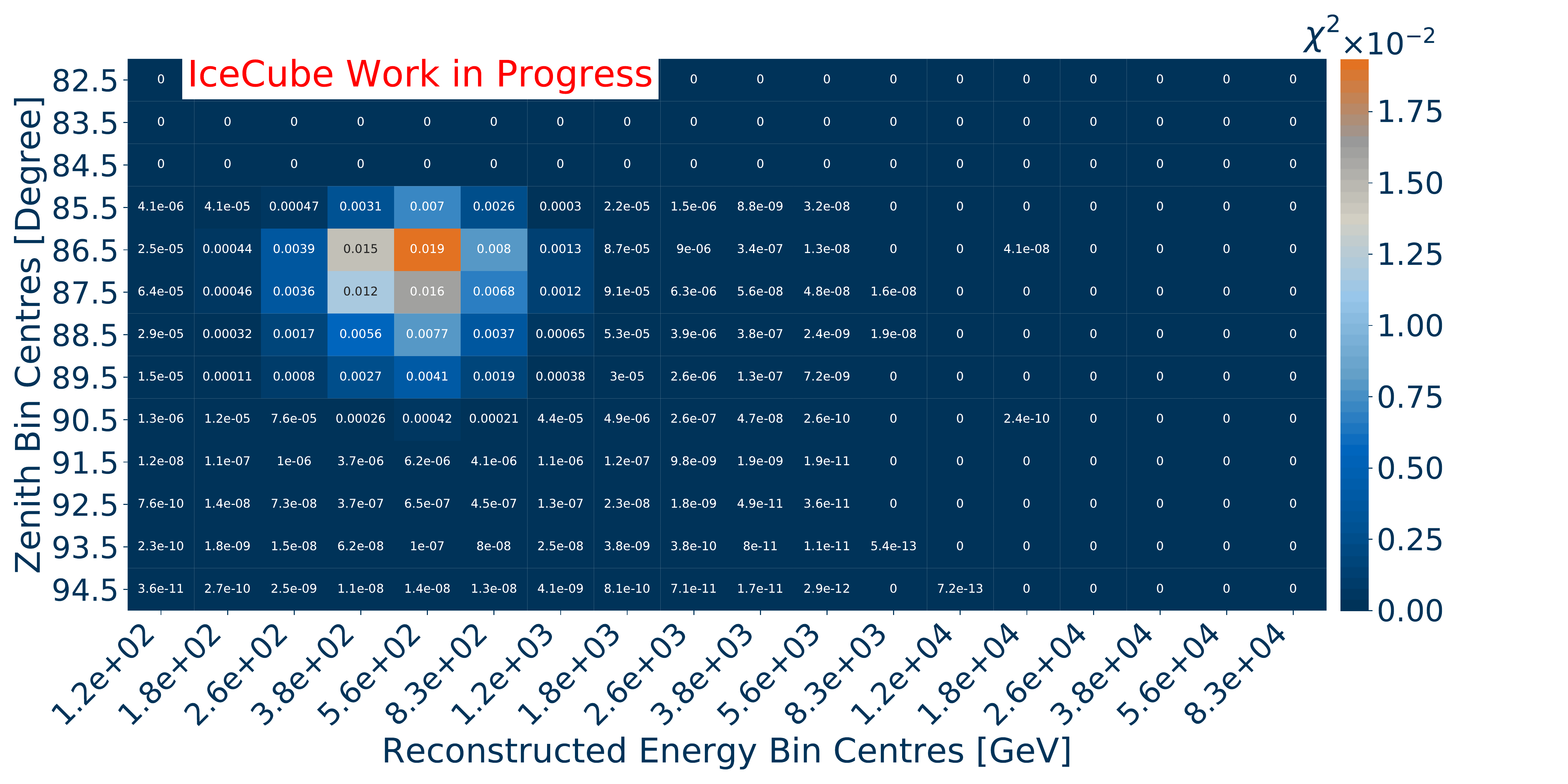}
    \caption{Signal over background $\chi^2$ data for a zenith energy bin grid.
    The data displayed is calculated according to \autoref{eq:chi2} for every bin of our region of interest.
    Values are for a stau mass of 100 GeV and one year of data.}
    \label{fig:chi2}
\end{figure}

A depiction of the $\chi^2$ values for the zenith/ energy grid assuming 100 GeV staus is shown in \autoref{fig:chi2}.
The mass dependence of $\chi^2$ is a result of the initial cross section that leads to a different flux at the detector.
Varying the number of stau events over background produces different $\chi^2$ maps for the different masses.

\paragraph{Sensitivity}
Another way of using the event rates ($R_{Signal},\; R_{Background}$) yields a sensitivity for the expected mass limit using an existing event selection.
We understand the rates as simulated averages for signal and background events.
Hence, for each bin we simulate $1 \times 10^5$ outcomes ($R_{Poisson}$ for signal and background with an underlying Poisson distribution.
With 
\begin{equation}
    \chi^2 = \frac{(R_{Poisson}-R_{Background})^2}{R_{Background}},
\end{equation}
we calculate the $\chi^2$ of each bin and experiment and finally build a sum over energy and zenith bins.
This leaves us with 10\textsuperscript{5} summed $\chi^2$ values for background and signal.

\begin{figure}
    \centering
    \includegraphics[width = .8\textwidth]{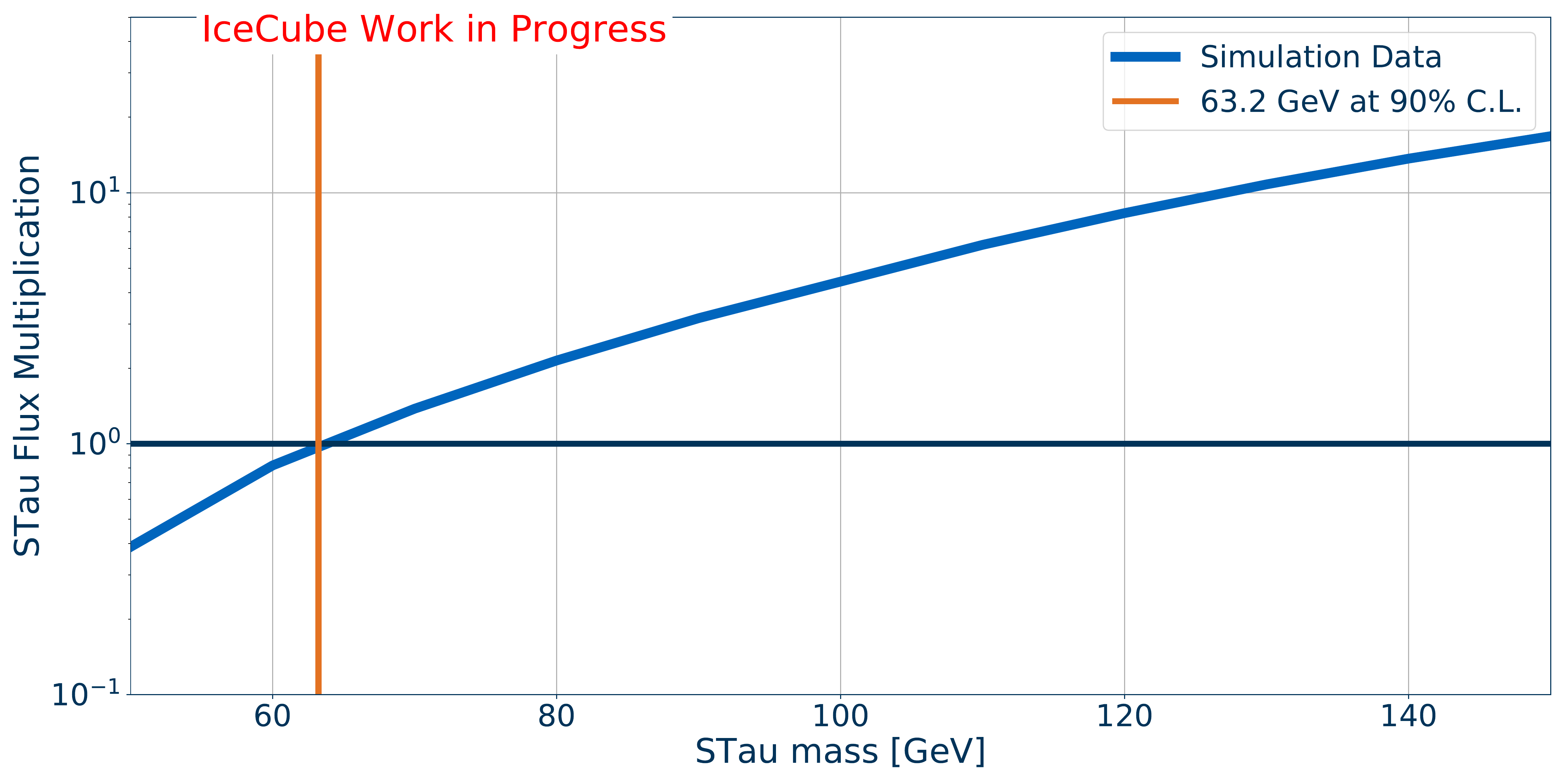}
    \caption{Sensitivity for expected stau mass limit.
    This figure shows the multiplication factor for simulated stau events necessary to achieve an exclusion of the corresponding stau mass.
    An exclusion is achieved if 90\% of the stau signals $\chi^2$ distribution lies above the median of the $\chi^2$ background distribution.}
    \label{fig:sensitivity}
\end{figure}

In our next steps we compare the median of all simulated $\chi^2$ values for the background to the tenth's percentile of the signals $\chi^2$ values.
Given a stau mass we can now find a multiplication factor for $R_{Signal}$ which we apply before simulating Poisson outcomes, so that both values agree.
This factor depends on the stau mass like shown in \autoref{fig:sensitivity}.
Our sensitivity then presents an expected limit by excluding all masses with a multiplication factor below one.

\paragraph{Outlook}
The current mass limit of the stau is determined by the collider experiment LHC to exclude masses < 430 GeV \cite{StaumassAaboud:2019trc}.
With our current event selection and neglecting systematic uncertainties, IceCube would be able to exclude stau masses < 63.2 GeV with 90\% confidence level.
Considering the state of our analysis, this sensitivity demonstrates the power of this type of study.
Using only existing and un-optimized tools like the event selection and and energy estimator and without any background rejection, this study misses the current limit by one order of magnitude.
However, improvements to our analysis are necessary to compete with the limits of collider experiments.

One of two main improvements involves an optimized event selection to be applied to the data.
The existing event selection of a different analysis works fine in our case for deriving a sensitivity.
However, for more advanced calculations it is unable to exploit the potential of the analysis.
An improved event selection targets a better effective area for staus as well as a reduction of background events.
It could incorporate a removal of the cut-off at 85° and open up the analysis to more sensitive zenith bins with a better signal to background ratio.

Further improvement of the sensitivity can be achieved with a better energy reconstruction.
A more precise reconstruction that fixes the bias for low-energy muons leads to fewer muon events in the sensitive energy bins.
In addition, it would produce a more sharp distribution of the stau signal.
If more stau events are reconstructed to the same energy bin, our signal to background ratio increases.

The effects of the named improvements are still to be determined but are expected to push the limit to LHC standards like proposed in \cite{Meighen-Berger:2020eun}.

\bibliographystyle{ICRC}
\bibliography{references}

\clearpage
\section*{Full Author List: IceCube Collaboration}




\scriptsize
\noindent
R. Abbasi$^{17}$,
M. Ackermann$^{59}$,
J. Adams$^{18}$,
J. A. Aguilar$^{12}$,
M. Ahlers$^{22}$,
M. Ahrens$^{50}$,
C. Alispach$^{28}$,
A. A. Alves Jr.$^{31}$,
N. M. Amin$^{42}$,
R. An$^{14}$
% ...
\\

\noindent
% ...
$^{12}$ Universit{\'e} Libre de Bruxelles, Science Faculty CP230, B-1050 Brussels, Belgium \\
% ...
$^{14}$ Department of Physics and Laboratory for Particle Physics and Cosmology, Harvard University, Cambridge, MA 02138, USA \\
% ...
$^{17}$ Department of Physics, Loyola University Chicago, Chicago, IL 60660, USA \\
$^{18}$ Dept. of Physics and Astronomy, University of Canterbury, Private Bag 4800, Christchurch, New Zealand \\
% ...
$^{22}$ Niels Bohr Institute, University of Copenhagen, DK-2100 Copenhagen, Denmark \\
% ...
$^{28}$ D{\'e}partement de physique nucl{\'e}aire et corpusculaire, Universit{\'e} de Gen{\`e}ve, CH-1211 Gen{\`e}ve, Switzerland \\
% ...
$^{31}$ Karlsruhe Institute of Technology, Institute for Astroparticle Physics, D-76021 Karlsruhe, Germany  \\
% ...
$^{42}$ Bartol Research Institute and Dept. of Physics and Astronomy, University of Delaware, Newark, DE 19716, USA \\
% ...
$^{50}$ Oskar Klein Centre and Dept. of Physics, Stockholm University, SE-10691 Stockholm, Sweden \\
% ...
$^{59}$ DESY, D-15738 Zeuthen, Germany \\
% ...

\subsection*{Acknowledgements}

\noindent
USA {\textendash} U.S. National Science Foundation-Office of Polar Programs,
U.S. National Science Foundation-Physics Division,
U.S. National Science Foundation-EPSCoR,
Wisconsin Alumni Research Foundation,
Center for High Throughput Computing (CHTC) at the University of Wisconsin{\textendash}Madison,
Open Science Grid (OSG),
Extreme Science and Engineering Discovery Environment (XSEDE),
Frontera computing project at the Texas Advanced Computing Center,
U.S. Department of Energy-National Energy Research Scientific Computing Center,
Particle astrophysics research computing center at the University of Maryland,
Institute for Cyber-Enabled Research at Michigan State University,
and Astroparticle physics computational facility at Marquette University;
Belgium {\textendash} Funds for Scientific Research (FRS-FNRS and FWO),
FWO Odysseus and Big Science programmes,
and Belgian Federal Science Policy Office (Belspo);
Germany {\textendash} Bundesministerium f{\"u}r Bildung und Forschung (BMBF),
Deutsche Forschungsgemeinschaft (DFG),
Helmholtz Alliance for Astroparticle Physics (HAP),
Initiative and Networking Fund of the Helmholtz Association,
Deutsches Elektronen Synchrotron (DESY),
and High Performance Computing cluster of the RWTH Aachen;
Sweden {\textendash} Swedish Research Council,
Swedish Polar Research Secretariat,
Swedish National Infrastructure for Computing (SNIC),
and Knut and Alice Wallenberg Foundation;
Australia {\textendash} Australian Research Council;
Canada {\textendash} Natural Sciences and Engineering Research Council of Canada,
Calcul Qu{\'e}bec, Compute Ontario, Canada Foundation for Innovation, WestGrid, and Compute Canada;
Denmark {\textendash} Villum Fonden and Carlsberg Foundation;
New Zealand {\textendash} Marsden Fund;
Japan {\textendash} Japan Society for Promotion of Science (JSPS)
and Institute for Global Prominent Research (IGPR) of Chiba University;
Korea {\textendash} National Research Foundation of Korea (NRF);
Switzerland {\textendash} Swiss National Science Foundation (SNSF);
United Kingdom {\textendash} Department of Physics, University of Oxford.

\end{document}